\documentclass[11pt]{article}

\usepackage{amsmath,amssymb,amsthm,mathtools,bm}
\usepackage{setspace,color,enumitem,url,booktabs,soul,multirow}

\usepackage{authblk}
\author[1,2]{\emph{Li-Chun Zhang}}
\affil[1]{\emph{Statistisk sentralbyrå, Norway}}
\affil[2]{\emph{University of Southampton, UK (L.Zhang@soton.ac.uk)}}
\title{Graph spatial sampling}
\date{}
  
\begin{document}

\maketitle

\begin{abstract} We develop lagged Metropolis-Hastings walk for sampling from simple undirected graphs according to given stationary sampling probabilities. It is explained how the technique can be applied together with designed graphs for sampling of units-in-space. We illustrate that the proposed graph spatial sampling approach can be more flexible for improving the design efficiency compared to the existing spatial sampling methods.   
\end{abstract}


\section{Introduction}

Denote by $U = \{ 1, ..., N\}$ the population of units-in-space, or simply \emph{units}. By \emph{graph spatial sampling (GSS)}, one would first \emph{design} a graph $G = (U, A)$ and then sample from $G$ -- hence its node set $U$ -- by graph sampling methods (Zhang, 2022; Zhang and Patone, 2017). The key idea is to sensibly introduce the edge set $A$, for which we consider only simple undirected graphs in this paper, in order to achieve certain desired spatial properties.  

\begin{figure}[ht]
\centering 
\includegraphics[scale=0.9]{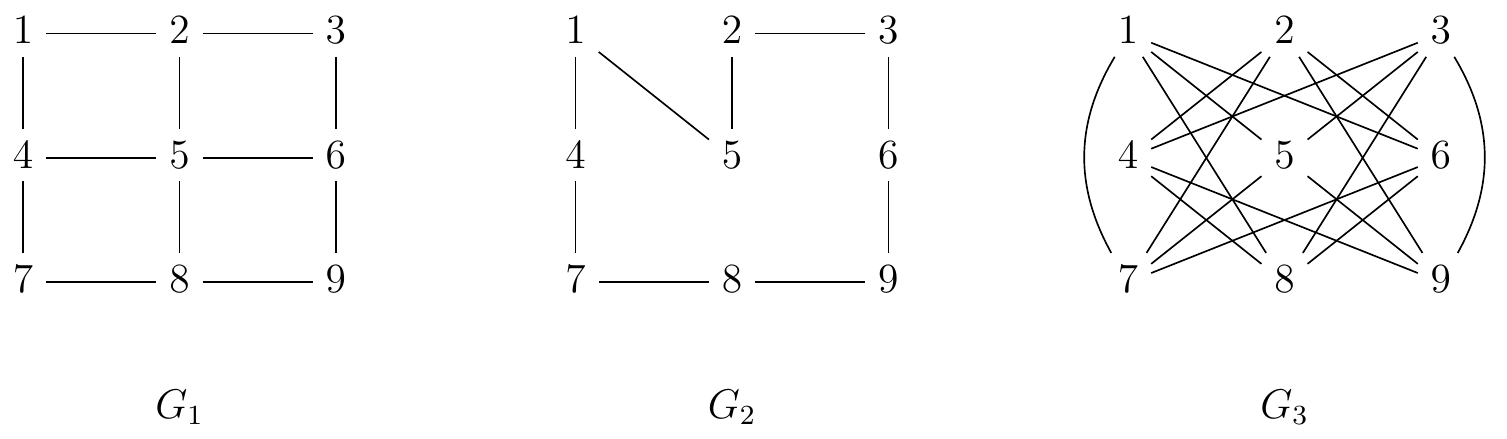} 
\caption{$G_1$, adjacency among all contiguous units; $G_2$, circular adjacency mostly among contiguous units; $G_3$, all contiguous units are non-adjacent} \label{fig:GSS}
\end{figure}

For instance, many spatial sampling methods aim to drastically reduce the chance of sampling contiguous (or nearby) units, compared to directly sampling from $U$ by non-spatial methods. To illustrate the idea in terms of GSS, three graphs are given in Figure \ref{fig:GSS} for sampling 2 out of 9 spatial units given as the nodes in $G$, where the edges defining the adjacency among the nodes are introduced in various ways. Depending on a chosen graph, one can employ different means for reducing the chance of selecting two contiguous units. 

First, in the popular generalised random tessellation stratified (GRTS) method (Stevens and Olsen, 2004), systematic sampling is applied to  the units in $U$ arranged on a path of $N-1$ edges which is a special graph $G = (U, A)$. More generally, the nodes can be arranged in a circle, as in the graph $G_2$ in Figure \ref{fig:GSS}. One can select a systematic sample of size 2 along the circle either clockwise or anti-clockwise. One can obtain an $(N-1)$-path for GRTS design by deleting one edge from $G_2$; however, it would then be impossible to select a systematic sample that is always of the size 2. Thus, the approach of GSS encompasses GRTS sampling. 

Moreover, instead of tessellation, one can also consider the graph $G_3$ in Figure \ref{fig:GSS}, where none of the contiguous units are adjacent. As it will be explained later, one can apply random walk without backtracking in $G_3$ and take as the sample the nodes visited by two successive steps of the walk at equilibrium, which are never contiguous. 

The local pivotal method (LPM) by Grafström et al. (2012) is another popular spatial sampling method, which can be applied to the graph $G_1$ in Figure \ref{fig:GSS}, where the contiguous units (adjacent in $G_1$) constitute the nearest units. As will be shown later, the LPM greatly reduces the chance of selecting an adjacent pair of nodes in $G_1$, compared to random sampling from $U$ directly; whereas GSS from $G_3$ avoids this altogether as described above. 

Notice that all the designs above are immeasurable. The GSS design using $G_3$ can be made measurable by allowing `random jumps' in addition, as will be explained later. However,  contiguous units can then be selected, e.g. by a random jump from 1 to 2. Generally, measurability is not considered a priority in spatial sampling for the sake of improved efficiency, but it does create a problem for variance estimation when the sampling is without replacement.

Below, we first develop a general technique of walk sampling from graphs in Section \ref{technique}, which achieves the desired sampling probabilities. In Section \ref{method}, we explain and illustrate how graph sampling can provide a more flexible approach to spatial sampling, compared to the existing popular methods. A discussion of some future topics is given in Section \ref{discussion}.

\section{Lagged Metropolis-Hastings walk} \label{technique}

Given $G = (U, A)$, let  $a_{ij} =1$ if $(ij)\in A$ and 0 otherwise. Let $d_i = \sum_{j\in U} a_{ij}$ be the degree of node $i$. We assume $d_i \geq 2$ for all the nodes in $G$ and there are no loops, such as the case with all the graphs in Figure \ref{fig:GSS}. At discrete time step $t>1$, let $X_t$ denote the state (i.e. the current node) of a \textit{lagged Metropolis-Hastings walk (LMHW)} in $G$, where $(X_0, X_1)$ are the two initial states, and the LMHW transition probability is given by 
\begin{align}
p_{(ih)j} \coloneqq \Pr(X_{t+1} =j \mid X_t =h, X_{t-1} =i) & = \frac{r u_j }{d_h +r} 
+ \mathbb{I}(j=i) \frac{w a_{hj}}{d_h +r} \min\Big( \frac{u_j}{u_h}, 1 \Big)  \notag \\
& \hspace{-20mm} + \mathbb{I}(j\neq i) \frac{d_h- w a_{ih}}{d_h +r} \big( \frac{a_{hj}}{d_h - a_{ih}} \big) \min\Big(\frac{u_j}{u_h}, 1\Big) 
\label{LMHW}
\end{align}
where $\bm{u} = (u_1, ..., u_N)$ is a positive \emph{preference vector} satisfying $\sum_{i\in U} u_i = 1$. That is, 
the walk either jumps randomly from $X_t = h$ to any node $j$ (in $U$) with the probability $u_j r/(d_h + r)$, or it moves to an adjacent node $j$ with the probability $d_h/(d_h +r)$. In the latter case, it can either backtrack to the \emph{previous}  $X_{t-1} =i$ (if adjacent) with a probability regulated by $w$ or move to \emph{another} adjacent node, both of which are subject to a Metropolis-Hastings (MH) acceptance mechanism, hence the term LMHW. There would be no backtracking under LMHW if $w=0$, and random jumps are disallowed if $r=0$ as long as $G$ is connected.

The LMHW \eqref{LMHW} generalises the lagged random walk (LRW) proposed by Zhang (2021) where $u_i \equiv 1$, i.e. without the MH  mechanism. Moreover, in the case of $w=1$, the LRW reduces to targeted random walk (TRW) of Avrachenkov et al. (2010), under which there is no difference between the previous state $X_{t-1}$ (if adjacent) and the other nodes adjacent to $X_t$. Thompson (2006) considers random walk (not lagged) subject to MH acceptance mechanism. 

The process $\{ X_t : t\geq 0\}$ is non-Markovian if $w<1$. Let $\bm{x}_t = (X_{t-1}, X_t)$ for $t\geq 1$. Given any initial $\bm{x}_1 = (X_0, X_1)$, LMHW \eqref{LMHW} generates a Markov chain $\{ \bm{x}_t : t\geq 1\}$, since
\begin{equation} \label{markov}
\Pr(\bm{x}_{t+1} | \bm{x}_t, ..., \bm{x}_1) = \Pr(\bm{x}_{t+1} | \bm{x}_t) = \Pr(X_{t+1} | \bm{x}_t)
\end{equation}
It is irreducible, if $G$ is connected or if random jumps are allowed generally, such that there exists a unique stationary distribution, $\Pr\big(\bm{x}_t = (h,j)\big)$, which is given by
\begin{equation} \label{px}
 p_{(hj)} = \sum_{i\in U} p_{(ih)} p_{(ih)j}
\end{equation}
A unique stationary distribution of $X_t$ follows, which satisfies the \emph{mixed} equation
\begin{equation} \label{mix}
p_h \coloneqq \Pr(X_t = h) = \sum_{i\in U}  \Pr\big( \bm{x}_t = (i,h)\big) 
= p_h p_{hh} + \sum_{\substack{\bm{x} = (i,h)\\ i\in \nu_h}} p_{\bm{x}} + \sum_{\substack{i\not \in \nu_h\\ i\neq h}}  \frac{p_i r u_h}{d_i +r} 
\end{equation}
where $p_{hh} \coloneqq \Pr(X_t = h | X_{t-1}=h)$ at equilibrium, and $\nu_h = \{ i : a_{ih} =1, i\in U\}$ is the \emph{neighbourhood} of $h$ (containing its adjacent nodes), and a transition from $h$ to any node outside $\nu_h$ can only be accomplished by a random jump. Notice that $\bm{x}_t = (h,h)$ for $t>1$ is possible if a random jump from $h$ lands on $h$ itself, or if a proposed move to an adjacent node is rejected. Appendix \ref{proof} gives a proof that the stationary probability is given by
\begin{equation} \label{p}
p_h \propto (d_h + r) u_h
\end{equation}
We have $p_h = \pi_h/n$ if $(d_h + r) u_h \propto \pi_h$ for all $h\in U$, where $\{ \pi_i : i\in U\}$ are the given sample inclusion probabilities by $\pi$ps sampling without replacement from $U$, where $\sum_{i\in U} \pi_i = n$.

\section{GSS by LMHW} \label{method}

\subsection{Equal-probability spatial sampling without replacement}

\emph{Equal-probability spatial sampling without replacement (EpSSWoR)} of sample size $n$ has the same sample inclusion probability $n/N$ as simple random sampling without replacement (SRSWoR) from $U$ directly, but the second-order inclusion probability of SRSWoR (which is the same for any pair of distinct units) can be modified to achieve desired spatial properties. 

LMHW can yield a GSS method for EpSSWoR. To ensure sampling without replacement, it is necessary to set $(r,w) = (0,0)$. In addition, to achieve equal probability \eqref{p} at equilibrium and to remove the possibility of rejecting any proposed transition to an adjacent node, devise a connected \emph{2-regular} graph $G$, where $d_i \equiv 2$, and set $u_i \equiv 1$. Now that $d_i \equiv 2$ and there is no backtracking, $X_{t+1}$ must be an adjacent node to $X_t$ which is not visited by the walk in the previous $N-1$ steps. It follows that any $n$-sequence of states $(X_t, X_{t+1}, ..., X_{t+n-1})$ is a sample of $n$ distinct units from $U$, where $p_h \equiv 1/N$ for any node $h$ in the sequence. 

\subsubsection{Illustration}

Now, one can devise the 2-regular graph $G$ according to the desirable spatial sampling properties. Suppose one would like to reduce the probability of selecting contiguous units, denoted by $\xi$. Consider below the spatial population $U$ in Figure \ref{fig:GSS} for an illustration. 

First, let the sample size be 2. There are 12 contiguous pairs (as can be seen in $G_1$) out of 36 possible pairs of distinct units, such that $\xi = 1/3$ under SRSWoR from $U$ directly. Simulations of the LPM1 version of LPM  (Grafström et al., 2012) from $G_1$ yields $\xi = 0.116$. The GRTS method cannot ensure the sample size is always 2 in this case. For GSS by clockwise systematic sampling from $G_2$ in Figure \ref{fig:GSS}, there are 9 systematic samples of size 2, i.e. $\{1,6\}$, $\{5,9\}$, ..., $\{7,2\}$ and $\{4,3\}$, where only $\{8,5\}$ contains contiguous units, such that $\xi =1/9 = 0.111$. 

\begin{figure}[ht]
\centering 
\includegraphics[scale=0.9]{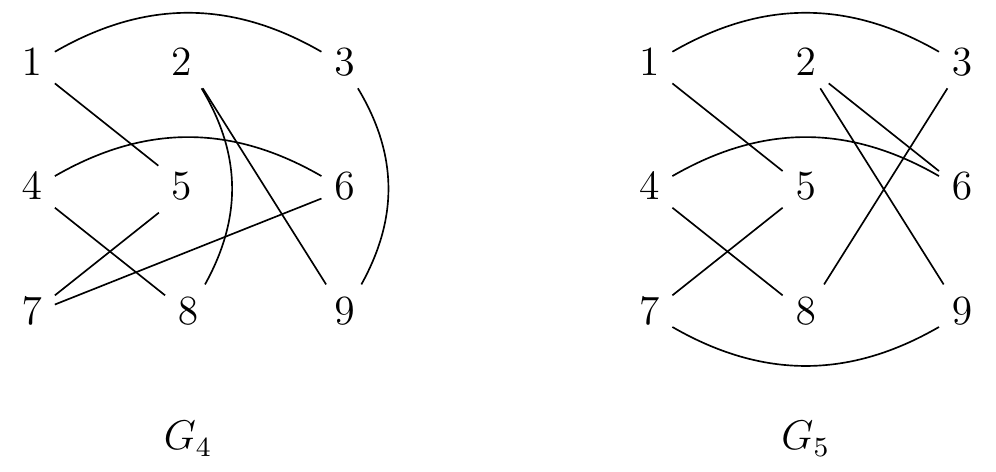} 
\caption{Two 2-regular graphs for EpSSWoR from 9 units-in-space} \label{fig:2G}
\end{figure}

Meanwhile, for EpSSWoR by LMHW \eqref{LMHW}, one can use a 2-regular graph that does not contain any edge connecting two contiguous units in $U$. There are many such graphs, two of which are shown in Figure \ref{fig:2G}. Using such a 2-regular graph, we obtain $\xi = 0$ by construction.

Next, let the sample size be 3. There are $84$ distinct samples by SRSWoR from $U$, where 22 of them do not contain any contiguous units, such that $\xi =62/84 =0.738$. Simulations of the LPM1 from $G_1$ yield $\xi = 0.478$. There are 3 systematic samples by the GRTS method, because $N/n = 3$. For instance, let the 8-path be given by $(1,4,7,8,9,6,3,2,5)$, i.e. removing the edge between 1 and 5 in $G_2$, the three samples are $\{1,8,3\}$, $\{4,9,2\}$ and $\{7,6,5\}$, such that  $\xi = 1/3$. The same holds for GSS by systematic sampling from $G_2$ in Figure \ref{fig:GSS}.

Meanwhile, for EpSSWoR by LMHW from $G_4$ (Figure \ref{fig:2G}), there are 9 distinct samples, where contiguous units are present in  the four samples $\{3,9,2\}$, $\{9,2,8\}$, $\{4,6,7\}$ and $\{6,7,5\}$, such that $\xi = 4/9 = 0.444$. However, suppose one instead adopts $G_5$ in Figure \ref{fig:2G}, then only the sample $\{6,2,9\}$ contains contiguous units, such that $\xi = 1/9 = 0.111$.

\subsubsection{Implementation} \label{G2r}

\begin{figure}[ht]
\centering
\includegraphics[scale=0.5]{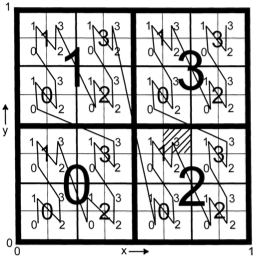} \hspace{20mm}
\includegraphics[scale=1.0]{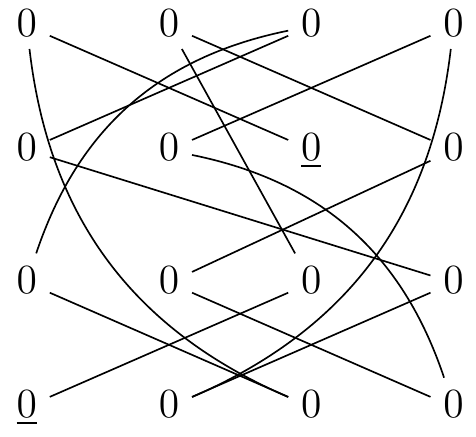} 
\caption{Recursive partition for GRTS design (left), 0-nodes in a 2-regular graph (right)} \label{fig:part}
\end{figure}

One can construct 2-regular graphs by means of the recursive partitions used for GRTS design. The example of Stevens and Olsen (2004) is given in Figure \ref{fig:part} (left), containing 64 units divided into 16 parts. Instead of connecting the nearby units as in the GRTS design, one can connect the more distant units, as illustrated for the 0-units in Figure \ref{fig:part} (right), which are non-contiguous due to the other units 1, 2, 3. The starting and end nodes are underlined in Figure \ref{fig:part}. Without loss of generality, suppose the one in the bottom-left corner is the end node. One can connect it to one of the 1-nodes that is not contiguous to the starting 0-node. Similarly for the other units 1, 2, 3. Finally, since the 3-nodes and 0-nodes are never contiguous here, connecting the end 3-node and the starting 0-node yields a non-contiguous 2-regular graph $G$.

Numerous 2-regular graphs can be devised like this; denote the collection of them by $\mathbb{G}$. For each $G$ in $\mathbb{G}$, let $\Omega_G$ contain the $N$ possible samples of the given size $n$. One can either calculate or simulate some \emph{design measure} over $\Omega_G$, denoted by $\tau_G$, such as $\xi$ above or the expected sample spatial balance (SSB)  measure of Stevens and Olsen (2004) or the sampling variance given $\{ y_i : i\in U\}$ generated by a suitable spatial population model. One can explore $(G, \tau_G)$ over $G\in \mathbb{G}$ and choose the graph $G$ that has the best design measure $\tau_G$. 

Given EpSSWoR by GGS-LMHW, a design-unbiased estimator of the population total 
\[
Y = \sum_{i\in U} y_i
\]
is the Horvitz-Thompson estimator. However, unbiased estimation of its sampling variance is impossible as long as the sampling design is immeasurable.

\subsection{Unequal-probability spatial sampling}

Let the graph $G$ for GSS be connected so that random jumps are unnecessary and set $r=0$. To further reduce the chance of selecting the same node more than once, set $w=0$ so that $\Pr(X_{t+1} = X_{t-1}) = 0$. Finally, set the preference vector $\bm{u}$, such that
\[
p_i = d_i u_i \eta = \frac{\pi_i}{n} \qquad\text{and}\qquad u_i = \frac{\pi_i}{n d_i \eta} \qquad\text{and}\qquad \eta = \frac{1}{n} \sum_{i\in U} \frac{\pi_i}{d_i} 
\] 
at equilibrium for any $i\in U$. Now, as long as $u_i$ is not a constant over $U$, one cannot avoid selecting some node more than once due to the rejections. Since the inclusion probability of any given node in $\{ X_{t+1}, X_{t+2}, ..., X_{t+m} \}$ becomes intractable as $m$ increases, we use the stationary sampling probabilities $p_i$ for unbiased estimation of the total $Y$.

\subsubsection{Based on $m$-sequence at equilibrium} 

Let $(X_{t+1}, X_{t+2}, ..., X_{t+m})$ be a sequence of $m$ states from the LMHW at equilibrium. An unbiased estimator of $Y$ can be given in various forms as
\begin{equation} \label{HH}
\hat{Y}_W = \frac{1}{m} \sum_{j=1}^m \frac{y_{X_{t+j}}}{p_{X_{t+j}}} 
= \frac{n}{m} \sum_{j=1}^{m} \frac{y_{X_{t+j}}}{\pi_{X_{t+j}}}  
= \frac{1}{m} \sum_{j=1}^{m} \sum_{i\in U} \frac{y_{X_{t+j}}}{p_{X_{t+j}}} \mathbb{I}(X_{t+j} = i)
\end{equation}
This includes $\pi_i \equiv n/N$ as a special case. We have $E(\hat{Y}_W | m) = Y$ because $\Pr\left[ \mathbb{I}(X_{t+j} = i)\right] = p_i$ given any time step $t+j$ under LMHW at equilibrium. Notice that the $m$ states in \eqref{HH} are not independent of each other, although the first expression of $\hat{Y}_W$ looks the same as the Hansen-Hurwitz estimator under sampling with replacement. 

Let $(X_{t+a}, ..., X_{t+b})$ be a subsequence of $(X_{t+1}, ..., X_{t+m})$, where $1\leq a \leq b\leq m$. It is said to a \emph{tie} of order $b-a+1$ for some node $h\in U$, denoted by $\kappa_{a,b}^h$, if 
\[
X_{t+a-1} = i \neq h,~ X_{t+a} = \cdots = X_{t+b} = h,~ X_{t+b+1} = j \neq h 
\]
Appendix \ref{tie} gives an unbiased estimator of $Y$ based on all the ties in an $m$-sequence. However, simulations suggest that it is typically less efficient than the simpler estimator \eqref{HH}.

\subsubsection{Illustration} \label{style}

Consider the following stylised examples of spatial populations $\bm{y}_U$ for Figure \ref{fig:GSS}:
\begin{align*}
& \left[ \begin{array}{ccc} 1 & 2 & 1\\ 2 & 3 & 2\\ 1 & 2 & 1 \end{array} \right] \qquad
\left[ \begin{array}{ccc} 3 & 2.5 & 2\\ 2.5 & 2 & 1.5\\ 2 & 1.5 & 1 \end{array} \right] \qquad
\left[ \begin{array}{ccc} 3 & 2 & 1\\ 2 & 1 & 2\\ 1 & 2 & 3 \end{array} \right] \qquad
\left[ \begin{array}{ccc} 3 & 2 & 3\\ 2 & 1 & 2\\ 3 & 2 & 3 \end{array} \right] \\
& \hspace{4mm} \text{Centre} \hspace{22mm} \text{Corner} \hspace{23mm} \text{Polar} \hspace{19mm} \text{Vortex}
\end{align*}
Let $\bm{\pi}_U$ be all equal if $\pi_5/\pi_i \equiv 1$, or unequal if $\pi_5/\pi_i \equiv 2$ or $\pi_5/\pi_i \equiv 0.5$, for all $i\neq 5$. We apply the LPM1 (Grafström et al., 2012) to select a sample of size 2, as well as LMHW sampling from each of $G_1$ - $G_5$ with $m=2$ and $r=w=0$. Simulations  yield the relative efficiency (RE) of a given sampling method against SRSWoR with $n=2$.

\begin{table}[ht] 
\centering
\caption{RE of LPM1 ($n=2$) or GSS ($m=2$) from $G_1$ - $G_5$}
\begin{tabular}{lcrrrrrr} \toprule
$\bm{y}_U$ & $\pi_5/\pi_i$ & LPM1 & $G_1$ & $G_2$ & $G_3$ & $G_4$ & $G_5$ \\ \midrule 
\multirow{4}{*}{Centre} & 1 & 0.92 & 1.27 & 0.58 & 0.57 & 0.87 & 0.85 \\
& & & (0.18) &  & & & \\ \cmidrule{2-8} 
& 2 & 0.62 & 0.32 & 0.08 & 0.56 & 0.84 & 0.82 \\
& & & (0.20) & (0.1) & (0.1) & (0.1) & (0.1) \\ \midrule
Corner & 1 & 0.71 & 1.91 & 1.72 & 0.77 & 0.66 & 0.68 \\ \midrule
\multirow{3}{*}{Polar} & 1 & 1.04 & 1.40 & 0.88 & 0.89 & 0.65 & 0.67 \\ \cmidrule{2-8} 
& 0.5 & 0.81 & 1.08 & 0.96 & 0.81 & 0.66 & 0.67 \\ 
& & & (0.27) & (0.07) & (0.07) & (0.07) & (0.07) \\ \midrule
\multirow{3}{*}{Vortex} & 1 & 0.91 &1.27 & 0.58 & 0.58 & 0.88 & 0.84  \\ \cmidrule{2-8}
& 0.5 & 0.58 & 0.45 & 0.18 & 0.46 & 0.74 & 0.73 \\ \bottomrule
\end{tabular} \label{tab:RE2} \\
Note: Positive $\Pr(n=1)$ by GSS given in parentheses
\end{table}

The results in Table \ref{tab:RE2} are based on $10^4$ simulations of each sampling method given $\bm{y}_U$. It is possible here that $\Pr(n=1) = \sum_{h\in U} p_{(hh)} >0$ under LMHW sampling due to the rejected moves, where the probability depends only on $(G, \bm{\pi}_U)$ but not $\bm{y}_U$. Setting $\pi_5/\pi_i \equiv 2$ can only be plausible for the centre $\bm{y}_U$, similarly as setting $\pi_5/\pi_i \equiv 0.5$ for the polar or vortex $\bm{y}_U$. Given $\pi_5/\pi_i \equiv 2$ for the centre $\bm{y}_U$, the two GSS methods using $G_1$ or $G_2$ select mostly contiguous units, both of which are actually more efficient than the other methods that aim to avoid selecting contiguous units; similarly given $\pi_5/\pi_i \equiv 0.5$ for the vortex $\bm{y}_U$. This serves as a reminder not to treat any particular sample spatial balance property as a panacea for design efficiency, without taking into account the spatial distribution of $\bm{y}_U$.

For equal-probability sampling across the 4 populations, although the LPM1 improves upon SRSWoR except in one case, it is always dominated by some (or all) of the GSS methods using $G_3$ - $G_5$. Among these GSS methods, using $G_4$ or $G_5$ yields essentially the same RE here, using $G_3$ is more efficient for the centre and vortex $\bm{y}_U$ but not otherwise. It is thus important to consider different graph designs for different spatial distributions of $\bm{y}_U$.

\subsection{Comparison of designs by simulation}

Grafström et al. (2012) suggest the LPM can yield large gains over the GRTS method for populations with smooth spatial trends, particularly in their Example 5 with $400$ units evenly spread over the unit square and $\pi_i \equiv n/N$ for $i\in U$, where the $y$-values are given by
\[
\text{sinTrend:}\quad y(x_1, x_2) = 3(x_1 + x_2) + \sin\{ 6(x_1 + x_2) \}
\]
and $(x_1, x_2)$ are the coordinates. We consider also the four types of $\bm{y}_U$ in Section \ref{style} for this $U$, where $0.5 \leq y_i\leq 5$ for $i\in U$, which is about the same range as the sinTrend $\bm{y}_U$ above. 

\begin{table}[ht] 
\centering
\caption{RE and ESSB of LPM1, $G_6$SS or $G_7$SS}
\begin{tabular}{llcccccc} \toprule
 & & \multicolumn{5}{c}{RE} & \\ \cmidrule{3-7}
Sample & Method & sinTrend & Centre & Corner & Polar & Vortex & ESSB \\ \midrule 
\multirow{3}{*}{$n=16$} & LPM1 & 0.151 & 0.248 & 0.127 & 0.221 & 0.244 & 0.080 \\
                            & $G_6$SS & 0.561 & 0.025 & 0.801 & 0.060 & 0.025 & 0.079 \\ 
                            & $G_7$SS & 0.044 & 1.371 & 0.016 & 1.047 & 1.362 & 0.192 \\ \midrule
 \multirow{3}{*}{$n=32$} & LPM1 & 0.090 & 0.147 & 0.072 & 0.132 & 0.150 & 0.074 \\
                             & $G_6$SS & 0.925 & 0.020 & 1.288 & 0.077 & 0.020 & 0.111 \\ 
                             & $G_7$SS & 0.027 & 1.489 & 0.009 & 1.227 & 1.543 & 0.238 \\ \midrule
\multirow{3}{*}{$n=48$}  & LPM1 & 0.067 & 0.111 & 0.053 & 0.098 & 0.114 & 0.079 \\
                             & $G_6$SS & 1.138 & 0.015 & 1.595 & 0.085 & 0.015 & 0.154 \\ 
                             & $G_7$SS & 0.022 & 1.421 & 0.007 & 1.321 & 1.375 & 0.238 \\ \bottomrule
\end{tabular} \label{tab:RE20}
\end{table}

Two 2-regular graphs are used for GSS here. The graph $G_6$ follows the description in Section \ref{G2r} (Figure \ref{fig:part}), with the $4\times 4$-partition of $U$ and 25 nodes in each part. The graph $G_7$ uses the $2\times 2$-partition as follows. First, index each unit $(x_1, x_2)$ as $(r_1, r_2)$, where $r_1$ is the rank of $x_1$ and $r_2$ that of $x_2$. Next, each pair of units $(r_1, r_2)$ and $(20-r_1+1, 20-r_2+1)$ are made adjacent, for $r_1, r_2 = 1, ..., 20$, i.e. between top-left and bottom-right parts as well as between top-right and bottom-left parts. Finally, the units in the top-left and bottom-left parts are randomly paired to be adjacent, likewise for the top-right and bottom-right parts. 

Table \ref{tab:RE20} gives the RE-results (each by $10^4$ simulations) and the expected sample spatial balance (ESSB), where the sample size $n\in \{ 16, 32, 48\}$ as in Grafström et al. (2012). For any $n$, $G_6$SS improves greatly over LPM1 for the Centre and Vortex $\bm{y}_U$, whereas $G_7$SS does so for the Corner and sinTrend $\bm{y}_U$. For the Polar $\bm{y}_U$, the RE is seen to become closer between LPM1 and $G_6$SS as $n$ increases, while both are considerably more efficient than SRSWoR. Notice that, since the ESSB is a constant given $n$ here, whichever the spatial population $\bm{y}_U$, one cannot anticipate the design efficiency \emph{only} based on such a measure. 

There exists a trend along $x_1+x_2$ in both the Corner and sinTrend $\bm{y}_U$, apart from a sinus undulation in the latter. The results suggest that the merits of $G_7$SS vs. LPM for the sinTrend $\bm{y}_U$ can be anticipated based on the Corner $\bm{y}_U$. Due to the structural similarity between the Centre and Vortex $\bm{y}_U$, the merits of $G_6$SS vs. LPM for one population can be anticipated from that for the other. The results for the Polar $\bm{y}_U$ suggest there may be room for improving the graph design for GSS  as $n$ increases for this and similar spatial populations.

\section{Some future topics} \label{discussion}

Random walk has numerous applications (e.g. Masuda et al., 2017; Brin and Page, 1998). LMHW offers a more flexible technique, which allows one to choose the desired stationary probabilities via the preference vector $\bm{u}$ while controlling the probability of back-tracking by $w$. It can be considered for many problems beyond spatial sampling. 

Both the GRTS method and the LPM can be motivated from the perspective of improving the expected SSB compared to sampling from $U$ directly. GSS provides a flexible approach to accommodate the anticipated spatial distribution of $\bm{y}_U$ in addition. It encompasses the GRTS method and, as illustrated above, suitable graph designs can yield large gains over the LPM. To facilitate the practice of GSS, one should develop suitable graph design algorithms that scale as the population size increases, and investigate their properties for various typical spatial distributions of $\bm{y}_U$ in a more systematic manner. 

For spatial sampling without replacement from $U$, variance estimation does not admit a theoretical solution. For GSS that allows for repeated selection of a given unit by LMHW, one can initiate multiple independent walks, each yielding an unbiased estimator \eqref{HH} --- one can use the mean of them to estimate $Y$ and use the between-walk variance of them for unbiased variance estimation, which is a standard technique in MCMC.

\appendix
\section{Proof of \eqref{p}} \label{proof}

Under LMHW \eqref{LMHW}, balanced flows between $\bm{x}_t = (i,h)$ and $\bm{x}_{t+1} = (h,j)$ are the flows over $(X_{t-1}, X_t, X_{t+1})$ in either direction. To show the values $\{ p_h = d_h+r : h\in U\}$ satisfy the balanced flows at equilibrium, one needs to consider the following situations I - V. 

\paragraph{I.} $i=h=j$, which is balanced at equilibrium, where the probability of either flow (in the opposite order) is equal to 
$\Pr\big( X_{t+1} = h | \bm{x}_t = (h,h)\big)$.

\paragraph{II.} $i=j\neq h$ and $i \in \nu_h$, i.e. $(i=j)$ \textemdash $~h$. Since, since $i= j$, both the flows $(i,h,j)$ and $(j,h,i)$ are backtracking (in either direction), the probability of which is the same by \eqref{LMHW}, so that they are always balanced.

\paragraph{III.} $i\neq j \in \nu_h$, i.e. $i$ \textemdash $~h$ \textemdash $~j$. Since $i\neq j$, neither $(i,h,j)$ nor $(j,h,i)$ is backtracking, given which \eqref{LMHW} yields
\begin{gather*}
\sum_{i\in \nu_h }\sum_{\substack{j\in \nu_h\\ j\neq i}} \Pr\big(\bm{x}_t = (i,h) \big) 
\Big( \frac{d_h -w}{d_h -1} \Big) \frac{\min(u_j, u_h)}{(d_h +r) u_h} \\
\sum_{j\in \nu_h }\sum_{\substack{i\in \nu_h\\ i\neq j}} \Pr\big(\bm{x}_t = (j,h) \big) 
\Big( \frac{d_h -w}{d_h -1} \Big) \frac{\min(u_i, u_h)}{(d_h +r) u_h}
\end{gather*}
as the sums of probabilities in either direction, which are balanced by symmetry.  

\paragraph{IV.} $\{ i, j\}\not \in \nu_h$, including $i=h$ or $j=h$, i.e. 
\begin{center}
$i\quad h\quad j$ \qquad or\qquad $(i=h)\quad j$ \qquad or\qquad $i\quad (h=j)$ \\
\end{center}
In the first case, where $i,h,j$ are distinct, any flow $(i,h,j)$ and $(j,h,i)$ can only take place by random jumps, which are balanced on setting $p_i \equiv (d_i +r) u_i$, since
\[
p_i \frac{r}{d_i +r} u_h \frac{r}{d_h +r} u_j = p_j \frac{r}{d_j +r} u_h \frac{r}{d_h +r} u_i
\] 
For the other two cases, on noting $p_h p_{hh} = \Pr\big( \bm{x}_t = (h,h)\big)$ by definition and setting $p_h = (d_h +r) u_h$, we obtain 
\[
\sum_{j\not \in \nu_h} p_h p_{hh} \frac{r u_j}{d_h +r} = \sum_{i\not \in \nu_h} p_h p_{hh} \frac{r u_i}{d_h +r}
\]
where the left-hand side is the sum of probabilities in the 2nd case (i.e. $i=h$) in the direction $(i,h,j)$, and the right-hand side is the sum of probabilities in the 3rd case (i.e. $h=j$) in the opposite direction $(j,h,i)$. It follows that these two cases balance out each other.

\paragraph{V.} One of $(i,j)$, say, $i$ belongs to $\nu_h$ but not the other, including when $j=h$, i.e.
\begin{center}
$i$ \textemdash $~h\qquad j$ \qquad or\qquad $i$ \textemdash $~(h=j)$
\end{center}
A flow in the direction $(i,h,j)$ consists of two parts: (i) $\bm{x} = (i,h)$ where $i\in \nu_h$, and (ii) either a random jump from $h$ to any node outside of $\nu_h$ (including $h$) or a proposed move into $\nu_i$ is rejected. Summing the stationary probabilities of all such flows, we have
\[
\big( \sum_{\substack{\bm{x} = (i,h)\\ i\in \nu_h}} p_{\bm{x}} \big) 
\left( \frac{r}{d_h +r} \sum_{j\not\in \nu_h} u_j + \frac{\sum_{i\in \nu_h} u_h - \min(u_i, u_h)}{(d_h + r) u_h} \right)
\]
where the 2nd term in the parentheses corresponding to (ii) is the sum over all possible moves, including backtracking to $X_{t-1} = i$ and forwarding to $l \neq X_{t-1}$ for all $i\in \nu_h$, which is a constant of $w$. Meanwhile, a flow in the opposite direction $(j,h,i)$ consists of two parts: (a) $\bm{x} = (j,h)$ where $j\not \in \nu_h$, including $j=h$, and (b) a transition from $h$ to an adjacent node. Summing the stationary probabilities of all such flows, we have
\[
\big( p_h -  \sum_{\substack{\bm{x} = (i,h)\\ i\in \nu_h}} p_{\bm{x}} \big) 
\left( \frac{r}{d_h+r} \sum_{i\in \nu_h} u_i + \frac{\sum_{i\in \nu_h} \min(u_i, u_h)}{(d_h +r) u_h} \right) 
\coloneqq \big( p_h -  \sum_{\substack{\bm{x} = (i,h)\\ i\in \nu_h}} p_{\bm{x}} \big) \Delta_h
\]
where the expression in the first pair of parentheses corresponding to (a) is by definition. To balancing the two groups of flows, we require
\begin{equation} \label{tmp}
p_h \Delta_h = \big( \sum_{\substack{\bm{x} = (i,h)\\ i\in \nu_h}} p_{\bm{x}} \big) \left( \frac{r}{d_h + r} + \frac{d_h u_h}{(d_h +r) u_h} \right)
= \sum_{\substack{\bm{x} = (i,h)\\ i\in \nu_h}} p_{\bm{x}}
\end{equation}
since $\sum_{i\in U} u_i = 1$, where the left-hand side in \eqref{tmp} can now be rewritten as 
\begin{align*}
p_h \Big( 1 -p_{hh} -\frac{r}{d_h +r} \sum_{\substack{i\not\in \nu_h\\ i\neq h}} u_i \Big) 
= p_h - p_h p_{hh} - r u_h \sum_{\substack{i\not\in \nu_h\\ i\neq h}} u_j 
\end{align*}
and the expression in the parentheses is $1 -\sum_{j\not \in \nu_h} \Pr(X_{t+1} = j | X_t = h)$ by definition, whereas $p_h = (d_h +r) u_h$ is used to obtain the last term on the right-hand side, which is equal to the last term on the right-hand side of \eqref{mix} on setting $p_i \equiv (d_i + r) u_i$. In other words, setting $p_h \equiv (d_h +r) u_h$ reduces \eqref{tmp} to \eqref{mix}, by which all the flows are balanced. 

Thus, balanced flows are achieved in all the situations above. This  completes the proof.

\section{Estimation based on $n_m$ ties in $m$-sequence} \label{tie}

Let the $m$-sequence $(X_{t+1}, ..., X_{t+m})$ consist of $n_m$ ties, denoted by $H_m = \{ h_1, ..., h_{n_m} \}$, where $n_m$ is random and $1\leq n_m \leq m-2$. The stationary probability of a tie $\kappa_{a,b}^h$ is given by
\[
p_{(\kappa_{ab}^h)} = \begin{cases} 
\sum\limits_{\substack{i,j\in U\\ i\neq h, j\neq h}} p_{(ih)} p_{(ih)j} & \text{if } a=b \\
\sum\limits_{\substack{i\in U\\ i\neq h}} p_{(ih)} p_{(ih)h} p_{(hh)h}^{b-a-1} (1-p_{(hh)h}) & \text{if } a<b
\end{cases}
\]
where $p_{(ih)}$ is given by \eqref{px} and all the transition probabilities by \eqref{LMHW}. 
 Let $\delta_{ab} =1$ if $(X_{t+a}, ..., X_{t+b})$ is a tie, and 0 otherwise. The conditional probability of $\kappa_{ab}^h$ given $I_{ab} = 1$ is 
\[
\bar{p}_{(\kappa_{ab}^h)} = p_{(\kappa_{ab}^h)} / \sum_{g\in U} p_{(\kappa_{ab}^g)} 
\]
Provided $\bar{p}_{(\kappa_{ab}^h)} >0$ for all $h\in U$, an unbiased estimator of $Y$ based on $H_m$ can be given by
\[
\hat{Y}_H = \frac{1}{n_m} \sum_{h\in H_m} y_h / \bar{p}_{(\kappa_{ab}^h)}
\]
We set $w=0$ to reduce the chance of selecting the same node by LMHW; we can allow for a small positive $r$ to ensure $\bar{p}_{(\kappa_{ab}^h)} >0$ for any $a<b$. We have then
\[
p_{(hh)h} = \frac{r u_h}{d_h +r} + \frac{1}{d_h +r} \sum_{j\in \nu_h}  \{ 1 - \min\Big(\frac{u_j}{u_h}, 1\Big) \} 
= \frac{r u_h + d_h - A_h}{d_h +r}
\]
where $A_h = \sum_{j\in \nu_h} \min\big(u_j/u_h, 1\big)$, and for $i\neq h$, 
\[
p_{(ih)h} = \frac{r u_h}{d_h +r} + \begin{cases} \frac{d_h - A_h}{d_h +r} & \text{if } i\not \in \nu_h \\ 
\frac{d_h}{d_h +r} \big\{ 1- \frac{1}{d_h -1} \{ A_h - \min\big(\frac{u_i}{u_h}, 1\big) \} \big\} & \text{if } i\in \nu_h
\end{cases}
\]
and for $i\neq h$ and $j\neq h$,
\[
p_{(ih)j} = \frac{r u_j}{d_h +r} + \begin{cases} 0 & \text{if } j \not \in \nu_h \text{ or } i=j \in \nu_h \\
\frac{1}{d_h +r} \min\big(\frac{u_j}{u_h}, 1\big) & \text{if } i\not \in \nu_h, j\in \nu_h \\
\frac{1}{d_h +r}\, \frac{d_h}{d_h -1} \min\big(\frac{u_j}{u_h}, 1\big) & \text{if } i\neq j \in \nu_h 
\end{cases}
\]

\end{document}